\documentclass[
 aip,
 amsmath,amssymb,
 reprint,%
]{revtex4-1}

\usepackage{graphicx}% Include figure files
\usepackage{dcolumn}% Align table columns on decimal point
\usepackage{bm}% bold math
\usepackage[utf8]{inputenc}
\usepackage[T1]{fontenc}
\usepackage{mathptmx}
\usepackage{etoolbox}

\usepackage{hyperref}
\usepackage{siunitx}
\DeclareSIUnit\inch{inch}
\DeclareSIUnit\oersted{Oe}
\DeclareSIUnit\unitformula{f.u.}
\DeclareSIUnit\angstrom{\text {Å}}
\DeclareSIUnit\bar{bar}

\makeatletter
\def\@email#1#2{%
 \endgroup
 \patchcmd{\titleblock@produce}
  {\frontmatter@RRAPformat}
  {\frontmatter@RRAPformat{\produce@RRAP{*#1\href{mailto:#2}{#2}}}\frontmatter@RRAPformat}
  {}{}
}%
\makeatother
\begin{document}

\preprint{AIP/123-QED}

\title[Bulk-like magnetic properties in MBE-grown unstrained, antiferromagnetic CuMnSb]{Bulk-like magnetic properties in MBE-grown unstrained, antiferromagnetic CuMnSb}

\author{L.~Scheffler}
\affiliation{Physikalisches Institut (EP3), Universit\"at W\"urzburg, 97074 W\"urzburg, Germany}
\affiliation{Institute for Topological Insulators,  Universit\"at W\"urzburg, 97074 W\"urzburg, Germany}
\author{J.~Werther}
\affiliation{Physikalisches Institut (EP3), Universit\"at W\"urzburg, 97074 W\"urzburg, Germany}
\affiliation{Institute for Topological Insulators,  Universit\"at W\"urzburg, 97074 W\"urzburg, Germany}
\author{K.~Gas}
\affiliation{Institute of Physics, Polish Academy of Sciences, Aleja Lotnikow 32/46, PL-02668 Warsaw, Poland.}
\author{C.~Schumacher}
\affiliation{Physikalisches Institut (EP3), Universit\"at W\"urzburg, 97074 W\"urzburg, Germany}
\affiliation{Institute for Topological Insulators,  Universit\"at W\"urzburg, 97074 W\"urzburg, Germany}
\author{C.~Gould}
\affiliation{Physikalisches Institut (EP3), Universit\"at W\"urzburg, 97074 W\"urzburg, Germany}
\affiliation{Institute for Topological Insulators,  Universit\"at W\"urzburg, 97074 W\"urzburg, Germany}
\author{M.~Sawicki}
\affiliation{Institute of Physics, Polish Academy of Sciences, Aleja Lotnikow 32/46, PL-02668 Warsaw, Poland.}
\author{J.~Kleinlein}
\email{johannes.kleinlein@physik.uni-wuerzburg.de}
\affiliation{Physikalisches Institut (EP3), Universit\"at W\"urzburg, 97074 W\"urzburg, Germany}
\affiliation{Institute for Topological Insulators,  Universit\"at W\"urzburg, 97074 W\"urzburg, Germany}
\author{L.~W.~Molenkamp}
\affiliation{Physikalisches Institut (EP3), Universit\"at W\"urzburg, 97074 W\"urzburg, Germany}
\affiliation{Institute for Topological Insulators,  Universit\"at W\"urzburg, 97074 W\"urzburg, Germany}

\date{\today}% It is always \today, today,

\begin{abstract}
A detailed study of the influence of molecular beam epitaxial growth conditions on the structural and magnetic characteristics of CuMnSb films on lattice matched GaSb is presented.
For a set of nine 40~nm thick layers, the Mn and Sb fluxes are varied to produce material with different elemental compositions.
It is found that the layers grown under a relative Mn to Sb flux ratio of $\Phi_{\text{Mn}}$/$\Phi_{\text{Sb}}=1.24\pm0.02$ are closest to the stoichiometric composition, for which the N\'{e}el temperature ($T_\text{N}$) attains its maximum values.
Mn-related structural defects are believed to be the driving contribution to changes in the vertical lattice parameter.
Having established the optimum growth conditions, a second set of samples with CuMnSb layer thickness varied from 5 to 510~nm is fabricated.
We show that for sufficiently large thicknesses, the magnetic characteristics ($T_\text{N}\simeq\SI{62}{\kelvin}$, Curie-Weiss temperature $\Theta_\text{CW} = -100$~K) of the stoichiometric layers do correspond to the parameters reported for bulk samples.
On the other hand, we observe a reduction of $T_\text{N}$ as a function of the CuMnSb thickness for our thinnest layers.
All findings reported here are of particular relevance for studies aiming at the demonstration of N\'{e}el vector switching and detection in this noncentrosymmetric antiferromagnet, which have been recently proposed.
\end{abstract}

\maketitle

Heusler alloys play an important role in the steadily growing field of antiferromagnetic spintronics.\cite{Jungwirth2018}
Their advantage over other material systems is twofold:
(i) they do not contain critical raw materials, iridium for example;\cite{Hirohata2017}
(ii) they offer a wide tunability of their material parameters while preserving other important functionalities. \cite{Graf2011}
One such example in the half-Heusler family is antiferromagnetic CuMnSb.\cite{Endo1968}
Its low N\'{e}el temperature, $T_\text{N}\simeq\SI{60}{\kelvin}$, which is convenient for standard laboratory equipment, makes CuMnSb a particularly suitable model platform for studying effects near and above the N\'{e}el transition.
Moreover, CuMnSb is a globally noncentrosymmetric antiferromagnet (the magnetic sites have symmetry group -42$m$) for which antidamping spin-orbit torques can be used for  N\'{e}el vector switching.\cite{Zelezny:2017_PRB}
Additionally, a sizable nonlinear anomalous Hall effect was predicted in CuMnSb due to the existence of a large Berry curvature dipole.\cite{Shao2020} 
The Berry curvature is supposed to be strongly dependent on the antiferromagnetic ordering, potentially providing N\'{e}el vector detection through Hall measurements.
Furthermore, in combination with its ferromagnetic counterpart, the half metallic half-Heusler NiMnSb, epitaxial ferromagnetic/antiferromagnetic heterostructures could be realized.

The functionality of spintronic devices depends on both electrical and magnetic material parameters, so control of both is crucial.
Magnetic properties of thin layers of the related half-Heusler compound NiMnSb, such as the Gilbert damping, saturation magnetization and exchange stiffness, as well as the vertical lattice parameter, can be tuned by small changes in material composition.\cite{Duerrenfeld2015}
In addition, magnetic in-plane anisotropy can be controlled by tuning either the material composition \cite{Gerhard2014} or the film thickness.\cite{Koveshnikov2005}
The highly controllable growth conditions in a molecular beam epitaxy (MBE) process make this method particularly suitable for investigating the influence of composition (a deviation from the stoichiometry) on the material properties.

Although bulk CuMnSb has been studied for over 50 years\cite{Endo1968}, many aspects of its magnetic behavior are not yet fully understood.
Experiments yield wide ranges of values for the characteristic material parameters such as $T_\text{N}$ (\SIrange[]{50}{62}{\kelvin}), the Curie-Weiss temperature $\Theta_\text{CW}$
(-250 to -120 K) and the effective moment $\mu_{\text{eff}}$ (3.9 to 6.3~$\mu_{\text{B}}$/f.u.).\cite{Endo1968,Forster1968,Endo1970,Helmholdt1984,Boeuf2006,Regnat2018,Bandyopadhyay2018}
In 2018, Regnat \textit{et al.} revealed the presence of a canted antiferromagnetic phase at low temperatures in bulk CuMnSb crystals grown by optical float-zoning.\cite{Regnat2018}
On theoretical grounds, it has been suggested that the feature-rich magnetic phase diagram of the material may be due to point defects.\cite{Maca2016}

Molecular beam epitaxy of strained antiferromagnetic CuMnSb thin films on InAs (001) substrates was recently demonstrated by our group.\cite{Scheffler2020}
For these layers, magnetic studies revealed two major deviations in magnetic properties as compared to bulk material:
An increase in Curie-Weiss temperature attributed to a reduction in geometric frustration due to reduced symmetry caused by epitaxial strain, and an additional linear region in the temperature dependent inverse susceptibility near the N\'{e}el temperature.
The origin of this second effect remains unclear.

Contrary to InAs, GaSb substrates can provide a strain-free platform for CuMnSb epitaxy.
In this letter we investigate the optimal growth conditions for CuMnSb (\mbox{$a_{\text{CuMnSb}}=\SI{6.095}{\angstrom}$, ref.~\onlinecite{Regnat2018}}) films on nearly lattice matched GaSb (001) (\mbox{$a_{\text{GaSb}}=\SI{6.09593}{\angstrom}$, ref.~\onlinecite{Levinshtein1996}}) substrates.
We investigate the effect of variations in material composition on the fundamental material properties.
Furthermore, the strain-free growth on GaSb allows the epitaxy of CuMnSb films without thickness limit, which we demonstrate up to a thickness of \SI{510}{\nano\metre}.

\begin{figure}[tb]
	\centering
		\includegraphics{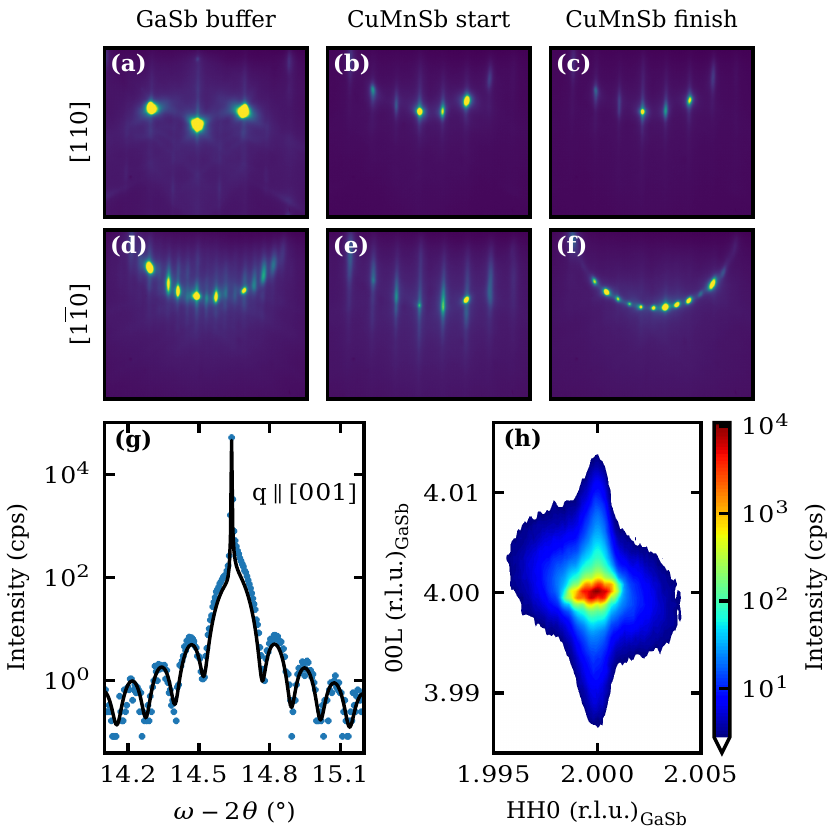}
	\caption{
        (a)-(f) Surface reconstructions observed by reflection high-energy electron diffraction during CuMnSb growth along the [110] and [1$\overline{\mbox{1}}$0] crystal directions.
        Within the first 2 minutes of CuMnSb growth, the $2\times5$ reconstruction of the GaSb buffer transforms into $2\times2$. After $\approx10$ minutes of growth, a $2\times4$ reconstruction develops, which remains unchanged until the end of the growth.
        (g) $\omega - 2\theta$ diffractogram around the symmetric (002) peak, together with a full dynamical simulation.
        (h) Reciprocal space map of the (224) diffraction peak.
    }
	\label{FIG:MBE}
\end{figure}

Separate growth chambers connected by an ultra-high vacuum transfer system are used to grow the individual layers.
To facilitate transport studies we use  non-conductive, low tellurium-doped epi-ready GaSb wafers.
In order to remove the natural oxide layer from the GaSb wafers, the substrates are heated to \SI{600}{\celsius} in an antimony flux with a beam equivalent pressure of $\text{BEP}_{\text{Sb}}=\SI{4e-6}{\milli\bar}$ provided by a single filament effusion cell.
After oxide desorption, a suitable 2D surface for epitaxial deposition of CuMnSb is achieved by growing a GaSb buffer layer.
For this purpose, the substrate is cooled to a temperature of \SI{530}{\celsius}, which is then maintained during buffer growth.
Growth starts by providing a Ga flux ($\text{BEP}_{\text{Ga}}=\SI{5.3e-7}{\milli\bar}$) using a dual filament effusion cell.
After 15 minutes of growth, the intended buffer thickness of \SI{150}{\nano\meter} is reached.
These flux levels, in combination with the selected growth temperature, result in a $1\times3$ surface reconstruction\cite{Lee1986,Bracker2000} as observed by reflection high-energy electron diffraction (RHEED).
Upon cooling, after the buffer growth is ended, the $1\times3$ reconstruction transforms into a $2\times5$ reconstruction at a temperature in the range of \SIrange{350}{400}{\celsius}, as shown in \autoref{FIG:MBE} (a) and (d).
At \SI{250}{\celsius}, the Sb flux used to stabilize the surface is turned off.

Subsequently, the sample is transferred to a dedicated Heusler growth chamber and the CuMnSb thin films are grown at \SI{250}{\celsius}  following the protocol given in ref.~\onlinecite{Scheffler2020}.
Before growth start, all three effusion cell shutters are opened for 5 minutes to stabilize the fluxes while the main shutter blanks all beams to the substrate.
Opening the main shutter initiates the growth of CuMnSb.
The GaSb $2\times5$ reconstruction [\autoref{FIG:MBE} (a) and (d)] transforms into a $2\times2$ reconstruction within the first 2 minutes of CuMnSb growth [\autoref{FIG:MBE} (b) and (e)].
For the sample we will later identify as the stoichiometric one, this $2\times2$ reconstruction transforms into the $2\times4$ one after 10 minutes of growth.
This reconstruction remains unchanged until the growth is terminated by closing the main shutter [\autoref{FIG:MBE} (c) and (f)].
The typical growth rate is 0.5~nm/min.
The growth is followed by magnetron sputtering of a 5~nm protective ruthenium cap in a separate chamber before the samples are removed from the ultra-high vacuum environment.

Structural characterization of the epitaxial layers is performed by high-resolution x-ray diffraction using a triple-axis diffractometer.
Full dynamical simulations \cite{Kriegner2013} of $\omega - 2\theta$ diffractograms around the symmetric (002) peak are used to extract lattice parameters and the layer thickness of the CuMnSb films.
\autoref{FIG:MBE} (g) shows this measurement together with the simulation trace for the stoichiometric sample.
The simulation yields a horizontal lattice parameter of \mbox{$a^{\parallel}_{\text{CuMnSb}} = a_{\text{GaSb}} = \SI{6.096}{\angstrom}$}, a vertical lattice parameter of \mbox{$a^{\perp}_{\text{CuMnSb}} = \SI{6.094}{\angstrom}$}, and a layer thickness of \SI{40}{\nano\metre}.
No signs of relaxation can be seen in the reciprocal space map of the (224) diffraction peak [\autoref{FIG:MBE} (h)].

To establish the exact growth conditions for stoichiometric material, we grow a range of nine \SI{40}{\nano\metre} thick CuMnSb layers, under the same Cu flux of \mbox{$\text{BEP}_{\text{Cu}}=\SI{5.80e-9}{\milli\bar}$} but with various Mn and Sb fluxes.
The Mn flux $\text{BEP}_{\text{Mn}}$ is varied between \SIrange[range-phrase={{ and }}]{8.65e-9}{9.61e-9}{\milli\bar}, while the Sb flux $\text{BEP}_{\text{Sb}}$ is varied between \SIrange[range-phrase={{ and }}]{4.00e-8}{4.23e-8}{\milli\bar}.
The relative flux ratios of Mn and Sb for the individual samples are then given by:\cite{Wood1982}
\begin{equation}
    \frac{\Phi_{\text{Mn}}}{\Phi_{\text{Sb}}}=\frac{\text{BEP}_\text{Mn}}{\text{BEP}_\text{Sb}}\cdot\frac{\eta_\text{Sb}}{\eta_\text{Mn}}\cdot\sqrt{\frac{T_\text{Mn} \mu_\text{Sb}}{T_\text{Sb} \mu_\text{Mn}}}.
\end{equation}
Here, $\eta$ is the ionization efficiency, $T$ is the effusion cell temperature, and $\mu$ is the atomic mass of the corresponding element.
Given that both elements are supplied by single filament effusion cells, it is understood that Mn is evaporated as single atoms and antimony as $\text{Sb}_4$ molecules.

\begin{figure}[tb]
	\centering
		\includegraphics{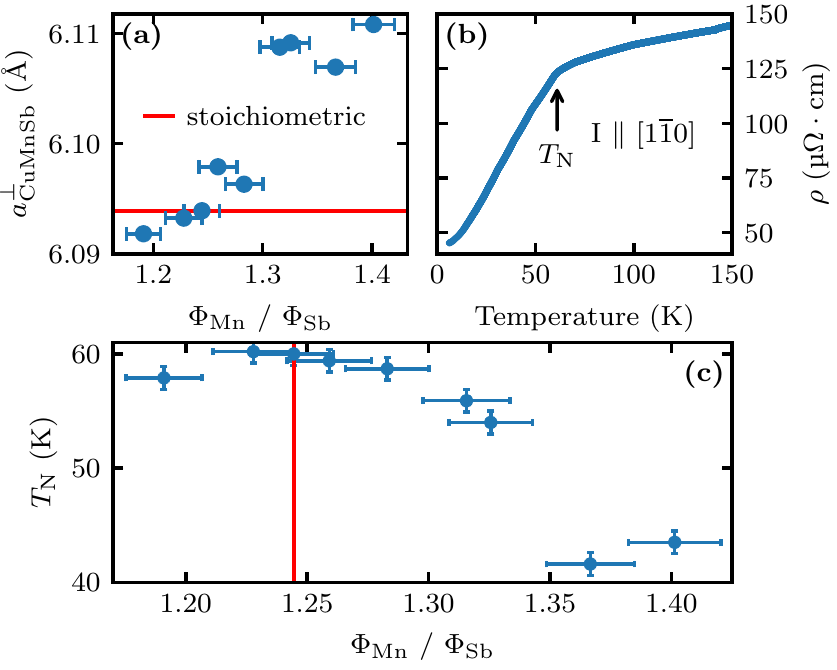}
	\caption{
        (a) Vertical lattice parameter of CuMnSb as a function of the $\Phi_{\text{Mn}}/\Phi_{\text{Sb}}$ flux ratio. The red line marks the calculated vertical lattice constant for stoichiometric CuMnSb grown on GaSb. 
        (b) Resistivity of the stoichiometric CuMnSb sample during warming with the current $I$ applied along the [1$\overline{\mbox{1}}$0] crystal direction.
        A clear kink at \SI{60}{\kelvin} marks the magnetic transition, as indicated by the arrow.
        (c) N\'{e}el temperature of all nine CuMnSb samples as a function of the $\Phi_{\text{Mn}}/\Phi_{\text{Sb}}$ flux ratio. The red line marks the $\Phi_{\text{Mn}}/\Phi_{\text{Sb}}$ flux ratio leading to the calculated vertical lattice constant for stoichiometric CuMnSb grown on GaSb.
    }

	\label{FIG:STOICHIOMETRY}
\end{figure}

\autoref{FIG:STOICHIOMETRY}~(a) shows the dependence of the vertical lattice parameter $a^{\perp}_{\text{CuMnSb}}$ on the relative Mn/Sb flux ratio for these nine layers.
An increase of the Mn flux leads to an enhancement of $a^{\perp}_{\text{CuMnSb}}$, as observed previously for epitaxial NiMnSb films.\cite{Roy2000,Gerhard2014,Duerrenfeld2015}
This increase is as large as \SI{0.3}{\percent}.
According to ref.~\onlinecite{Maca2016} $\text{Mn}_\text{Cu}$ (Mn substituting Cu), $\text{Cu}_\text{Mn}$, and $\text{Mn}_\text{vac}$ (interstitials) should be the dominant defects responsible for the increase of $a$, since their formation energy is the lowest in CuMnSb.
Moreover, it has been suggested that such defects are responsible for an increase of $a$ in NiMnSb.\cite{Ekholm2010}
Consequently, it is likely that the variation of the Mn content is responsible for the variation of $a^{\perp}_{\text{CuMnSb}}$ observed here.

In order to determine which sample corresponds best to stoichiometric conditions, we compare the vertical lattice parameters obtained from the $\omega - 2\theta$ scans with the calculated lattice parameter for CuMnSb grown pseudomorphically on GaSb.
By using the cubic elastic tensor\cite{Palaz2017} and the bulk lattice parameter\cite{Regnat2018} of CuMnSb (determined by X-ray powder diffraction), we obtain the magnitude of the vertical lattice parameter for stoichiometric conditions as \mbox{$a^{\perp}_{\text{stoi}} = \SI{6.094}{\angstrom}$}.
The red line in \autoref{FIG:STOICHIOMETRY}~(a) marks this value.
The position of the red line in \autoref{FIG:STOICHIOMETRY}~(a) indicates that the sample grown with the relative flux ratio of $\Phi_{\text{Mn}}$/$\Phi_{\text{Sb}}=1.24\pm0.02$ matches $a^{\perp}_{\text{stoi}}$ best.
Accordingly, we designate this sample as the stoichiometric one.

For the electrical characterization of these films, \mbox{$1080\times180$~\si{\micro\metre\squared}} sized Hall bars are fabricated by standard optical lithography.
The mesa is defined by argon dry etching followed by wet etching in a phosphoric-citric acid based solution.
To account for parasitic conductivities due to the Ru protective layer and the substrate, the measurements are repeated with a reference sample. It consists of the same layers as the CuMnSb sample, but lacks the CuMnSb layer. The conductivity of the reference sample is then subtracted from the measurements following the process described in ref.~\onlinecite{Scheffler2020}.

The temperature dependence of the resistivity $\rho(T)$ for the stoichiometric CuMnSb layer is given in \autoref{FIG:STOICHIOMETRY}~(b).
At around $T\simeq\SI{60}{\kelvin}$  a clear kink indicating the suppression of spin disorder scattering, manifests the presence of a magnetic phase transition.\cite{Mott1964}
Since an equivalent sharp structure appears during our magnetic measurements (see below), we attribute this kink to the paramagnetic-antiferromagnetic phase transition and associate its $T$-position with $T_\text{N}$.
\autoref{FIG:STOICHIOMETRY}~(c) summarizes the $T_\text{N}$ of the various samples as a function of $\Phi_{\text{Mn}}/\Phi_{\text{Sb}}$.
The red line in \autoref{FIG:STOICHIOMETRY}~(c) marks the flux ratio where we expect the stoichiometric composition.
The highest values of $T_\text{N}$ (up to $T_\text{N}\simeq\SI{60}{\kelvin}$) are grouped around the stoichiometric growth conditions ($\Phi_{\text{Mn}}$/$\Phi_{\text{Sb}}=1.24$).
Leaving the stoichiometric composition, leads to a decrease of $T_\text{N}$, similar as reported for Co substitution in CuMnSb.\cite{Bandyopadhyay2018}
Moreover, the two samples with the highest $\Phi_{\text{Mn}}/\Phi_{\text{Sb}}$ exhibit spontaneous magnetization already at room temperature, suggesting the presence of ferromagnetic inclusions due to the high Mn concentration.

Having established the optimum growth conditions for the stoichiometric composition of CuMnSb epilayers, we turn to the investigation of the role of the thickness $t$ on the magnetic properties of CuMnSb.
To this end, seven additional CuMnSb layers with $t$  ranging from \SIrange[range-phrase={{ to }}]{5}{510}{\nano\metre} are grown with stoichiometric material composition, i.e. with $\Phi_{\text{Mn}}$/$\Phi_{\text{Sb}}=1.24$, as already established.

The magnetic measurements are performed in a commercial superconducting quantum interference device (SQUID) magnetometer.
The measurements follow the established protocols for investigating minute magnetic signals,\cite{Sawicki2011} modified to account for some shortcomings of commercial magnetometers built around superconducting magnets.\cite{Gas2019}
It is important to meticulously remove the metallic MBE glue from the backside of the samples, as its magnetic contribution would otherwise be of the same order\cite{Gas2021} as that expected from our thin films.
As with the transport measurements a reference sample grown without the CuMnSb layer is investigated.
The magnetic moment of the CuMnSb layer can then be determined quantitatively as follows:
\begin{equation}
    m_{\text{CuMnSb}}=\frac{m_S}{\gamma_S}-\frac{m_R}{\gamma_R}\cdot\frac{\mu_S}{\mu_R},
\end{equation}
where $m_{S(R)}$ is the measured magnetic moment, $\gamma_{S(R)}$ is a correction factor which scales the magnetometer output of a three-dimensional sample to its equivalent for a point dipole, and $\mu_{S(R)}$ is the mass of the sample (reference).
Importantly, for asymmetrical objects $\gamma$ depends on the orientation of the object with respect to the axis of the magnetometer.\cite{Stamenov2006,Sawicki2011}

\begin{figure}[tb]
	\centering
		\includegraphics{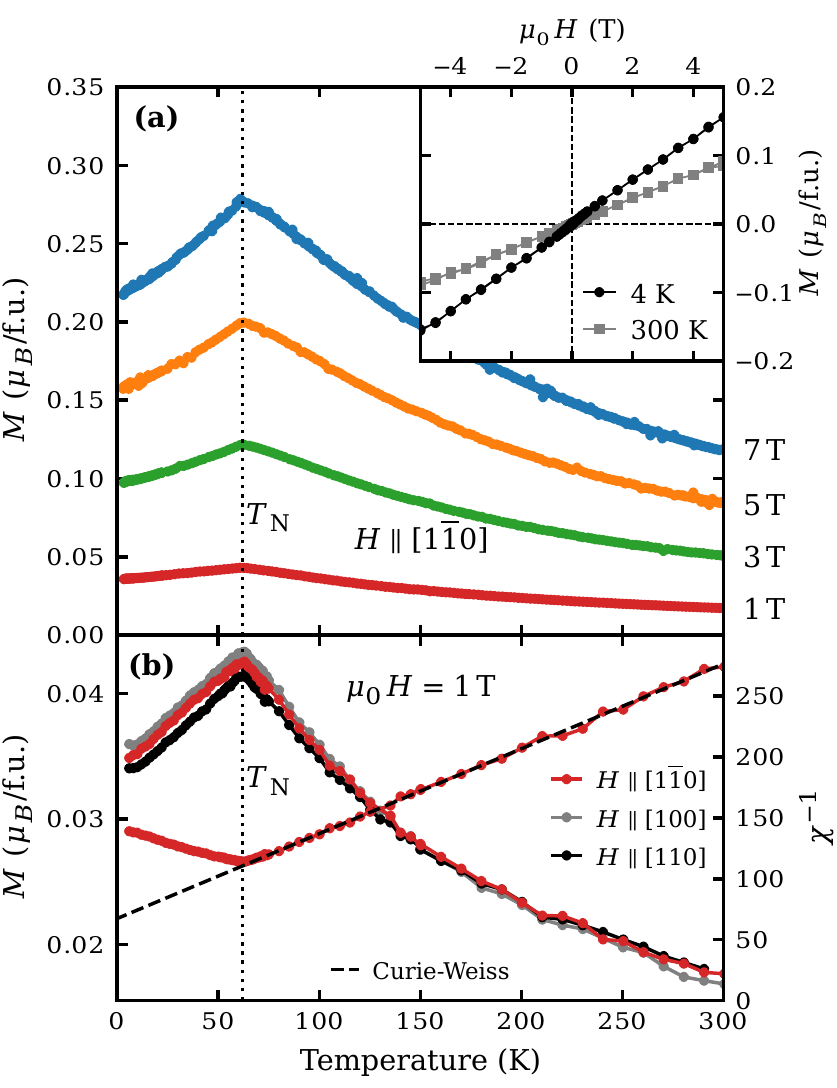}
	\caption{
        (a) Temperature dependence of magnetization $M(T)$ of a 510~nm thick CuMnSb layer taken at selected external magnetic fields $\mu_0 H=$ 1, 3, 5, and 7~T
        along the [1$\overline{\mbox{1}}$0] crystal direction.
        The vertical dotted line marks the N\'{e}el temperature $T_\text{N} = 62$~K.
        The linear response of $M$ with respect to $H$ is shown in the inset, where two $M(H)$ dependencies along [1$\overline{\mbox{1}}$0] measured in the aniferromagnetic phase at 4~K and in the paramagnetic phase at 300~K are shown.
        (b) Left axis: $M(T)$ of the same layer taken $\mu_0 H= 1$~T along the [1$\overline{\mbox{1}}$0], [100] and [110] crystal directions together with (right axis) the inverse of the magnetic susceptibility $\chi(T)^{-1}$ obtained from the $M(T)$ along [1$\overline{\mbox{1}}$0].
        The dashed line indicates the Curie-Weiss law.
    }
	\label{FIG:SQUID}
\end{figure}

First, we examine the magnetic properties of the thickest CuMnSb layer ($t=510$~nm), since its magnetic response is strongest.
\autoref{FIG:SQUID} summarizes our studies.
In \autoref{FIG:SQUID}~(a) the $T$-dependent magnetization curves, $M(T)$, obtained for $\mu_0 H= 1-7$~T, confirm the presence of the N\'{e}el transition at an $H$-independent $T_\text{N} = 62$~K.
Quantitatively, our results are identical to those reported for bulk samples,\cite{Regnat2018}
however the magnitudes of $M$ established here are about 44\% larger and the value of $T_\text{N} = 62$~K is about 7~K higher than the results reported there.
Importantly, our $M(T< T_\text{N})$ data do not indicate any features which could mark a tilting of the Mn spins from the $\langle 111 \rangle$ direction, i.e. from the expected orientation of Mn sublattices in this face-centered cubic structure.
It remains to be seen whether these experimental differences highlight a disparity between stoichiometric and a point-defect-affected material.

Since the magnetic response is linear in $H$ in the whole range of $H$ and $T$ as indicated in the inset in \autoref{FIG:SQUID}~(a), we
take a closer view  on $M(T)$ by examining only the $\mu_0 H= 1$~T case.
In \autoref{FIG:SQUID}~(b) we show \mbox{$M(T,\mu_0 H= 1$~T)} obtained for all major in-plane crystal directions, [1$\overline{\mbox{1}}$0], [100] and [110].
The results are found to be fairly isotropic, however, a very weak dependence on the orientation is observed below 100~K.
A similar behavior was also found in bulk CuMnSb.\cite{Regnat2018}
The right axis of this panel denotes the magnitude of the inverse of the magnetic susceptibility, $\chi(T)^{-1}$.
Above $T_\text{N}$ the inverse susceptibility follows a perfectly straight line.
Its abscissa and slope yield $\Theta_\text{CW} = -100$~K and $\mu_{\text{eff}} = 5.6~\mu_{\text{B}}$/f.u., respectively.
These values are in line with the previously reported values ($\Theta_\text{CW}$ and $\mu_{\text{eff}}$ of compressively strained CuMnSb/InAs are $-65$~K and $5.9~\mu_{\text{B}}$/f.u.\cite{Scheffler2020}, respectively), with $\mu_{\text{eff}}$ slightly exceeding the theoretically expected value of $4.9~\mu_{\text{B}}$/f.u..\cite{Jeong2005}

\begin{figure}[tb]
	\centering
		\includegraphics{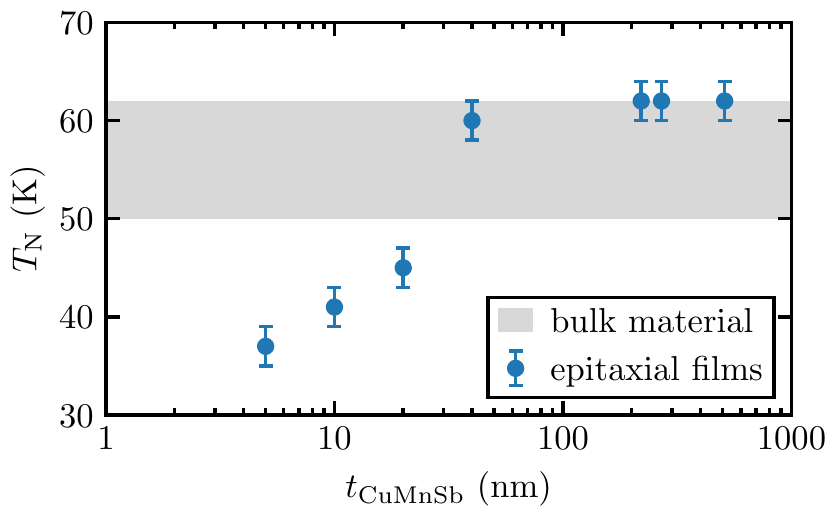}
	\caption{
        N\'{e}el temperature $T_\text{N}$ as a function of the CuMnSb layer thickness $t$.
        The range of values reported in the literature for bulk material is marked in gray.
    }
	\label{FIG:THICKNESS}
\end{figure}

Temperature dependent measurements like those presented in \autoref{FIG:SQUID}~(a) and (b) are used to determine the N\'{e}el temperature dependence on thickness for stoichiometric layers.
We plot the results  in \autoref{FIG:THICKNESS} (bullets).
For reference, the thick gray band at the background marks the range of N\'{e}el temperatures established for bulk CuMnSb.
The data obtained here indicate that layers with \mbox{$t\geq\SI{200}{\nano\metre}$} undergo the magnetic transition at $T_\text{N}=\SI{62}{\kelvin}$,
i.e. their N\'{e}el temperatures correspond to the highest values established previously\cite{Bandyopadhyay2018} for bulk CuMnSb.

On the other hand, $T_\text{N}$ starts to decrease rapidly below \SI{\sim 50}{\nano\metre}, and for the thinnest sample (\mbox{$t=\SI{5}{\nano\metre}$}) has dropped to \mbox{$T_\text{N}=\SI{37}{\kelvin}$}, potentially indicating a difference in surface and bulk contributions to the antiferromagnetic order.

In summary, we report the successful growth of CuMnSb layers on low tellurium-doped GaSb (001) substrates by molecular beam epitaxy.
The very small lattice mismatch allows epitaxial CuMnSb layers to be grown with unlimited thickness.
We find that the relative flux ratio of Mn and Sb, $\Phi_{\text{Mn}}$/$\Phi_{\text{Sb}}=1.24\pm0.02$, during the deposition leads to layers closest to the calculated vertical lattice constant for stoichiometric material composition.
Such layers exhibit maximum values of their N\'{e}el temperature ($T_\text{N} = 62$~K for the thickest layers studied here).
We postulate therefore that this particular value of $T_\text{N}$ may serve as a practical selection tool of CuMnSb samples with the stoichiometric composition and together with $\Theta_\text{CW} = -100$~K and $\mu_{\text{eff}} = 5.6~\mu_{\text{B}}$/f.u. serve as benchmarks for this material system.
Mn-related defects are believed to be the driving force for changes in the vertical lattice parameter when the flux ratio is changed.
From the thickness dependence studies it has been concluded that the epitaxial CuMnSb layers behave like bulk material for thicknesses above \SI{200}{\nano\metre}.
On the other hand, $T_\text{N}$ is significantly reduced in much thinner layers, dropping down to 37~K for the 5~nm thin sample.
The deposition method of high quality stoichiometric epitaxial layers of CuMnSb shown here is an important step towards practical spintronics functionalization of the noncentrosymmetric antiferromagnet.

\begin{acknowledgments}
We thank M.~Zipf and V.~Hock for technical assistance.
This work is funded by the Deutsche Forschungsgemeinschaft (DFG, German Research Foundation) - 397861849 and by the Free State of Bavaria (Institute for Topological Insulators) and the Deutsche Forschungsgemeinschaft (DFG, German Research Foundation) under Germany's Excellence Strategy-EXC2147 "ct.qmat" (project-id 390858490).
\end{acknowledgments}

\section*{Data Availability Statement}
The data that support the findings of this study are available from the corresponding author upon reasonable request.

\bibliography{literatureNew}% Produces the bibliography via BibTeX.

\end{document}